\documentclass[twoside,slac_one]{revtex4}
\usepackage{graphicx}
\usepackage{fancyhdr}
\usepackage{amsmath} % American Mathematics Society standards
\usepackage{bm}% bold math
\usepackage{amsxtra}
\usepackage{amssymb}
\usepackage{amsthm}
\usepackage{latexsym}
\usepackage{lscape}
\usepackage{url}
\begin{document}

\title{Muon Collider: Plans, Progress and Challenges}


\author{Ronald Lipton}
\affiliation{Fermi National Accelerator Laboratory,  Batavia, Il, USA}

\begin{abstract}
We in the physics community expect the LHC to uncover new physics in the 
next few years.  The character and energy scale of the new 
physics remain unclear, but it is likely that data from the LHC will need to be complemented 
by information from a lepton collider  which can provide for precise examination of new 
phenomena.  We describe the concept, accelerator design, and detector R\&D for 
a high energy Muon Collider as well as the challenges associated with the machine and its 
detector environment.
\end{abstract}

\maketitle

\thispagestyle{fancy}


%%%%%%%%%%%%%%%%%%%%%%%%%%%%%%%%%%
\section{Muon Colliders - Who Ordered That?}
The use of muons in a high energy collider appears to be a desperate measure. 
After all muons have a lifetime of $2.2 \mu s$ and will only survive for 
about 2000 turns in a 1.5 TeV storage ring. 
But muons also have distinct advantages as projectiles in a colliding beams accelerator.
They are pointlike, so one can adjust the center of mass energy of the collision 
precisely and study resonance structures and threshold effects in great detail. 
Secondly, they are 207 times more massive than the electron, meaning that muons
radiate $(m_\mu/m_e)^4$ times less than an electron traveling with he same radius of curvature and 
energy.  This means that a muon collider is a circular, rather than a linear, 
machine.  It also means that beamsthralung effects, radiation due to beam-beam 
interactions, would be much smaller in a muon collider than an $e^+e^-$ machine, allowing 
for precise beam constraints and energy measurements. Finally the mass-dependent coupling 
of the Higgs to the ($\mu^+\mu^-$) system is 40,000 times larger than the coupling to $e^+e^-$, making 
a muon collider an ideal candidate for direct study of s channel Higgs.

Muons in a storage ring are reused, each muon has $\approx$2000 chances to collide with the 
opposing beam before it decays.  This relaxes requirements on emittance and makes    
constraints on beam dimensions much more forgiving than a linear collider with its 
extremely tight beam focussing requirements. 
 Such a machine would also be compact, with machines up to 5 TeV fitting on the 
Fermilab site.  The machine is also likely to use significantly less power than a comparable 
electron collider. Initial calculations indicate the wall power needed for a 3 TeV Muon Collider would 
be 1/3 of that for a 3 TeV CLIC and 2/3 of that for a 0.5 TeV ILC (Table~\ref{example_table}). 
 A muon collider could scale to multi-TeV 
energies without excessive penalties in power or cost.

%%%%%%%%%%%%%%%%%%%%%%%%%%%%%%%%%%
\subsection{Accelerator Challenges}
The short muon lifetime means that everything must be done quickly.  Muons must be produced and collected, 
cooled, and re-accelerated rapidly. This forces many of the components to provide multiple functions, 
combining cooling,  acceleration, and focussing. The beam transport system must handle the radiation and heat load 
associated with electrons from muon decay. Detectors must be well-shielded from the bulk of the decay backgrounds.
The Muon Accelerator Program (MAP)~\cite{MAP} has been organize to study the many technical 
challenges associated with a Muon Collider.

\begin{figure*}[ht]
\centering
\includegraphics[width=135mm]{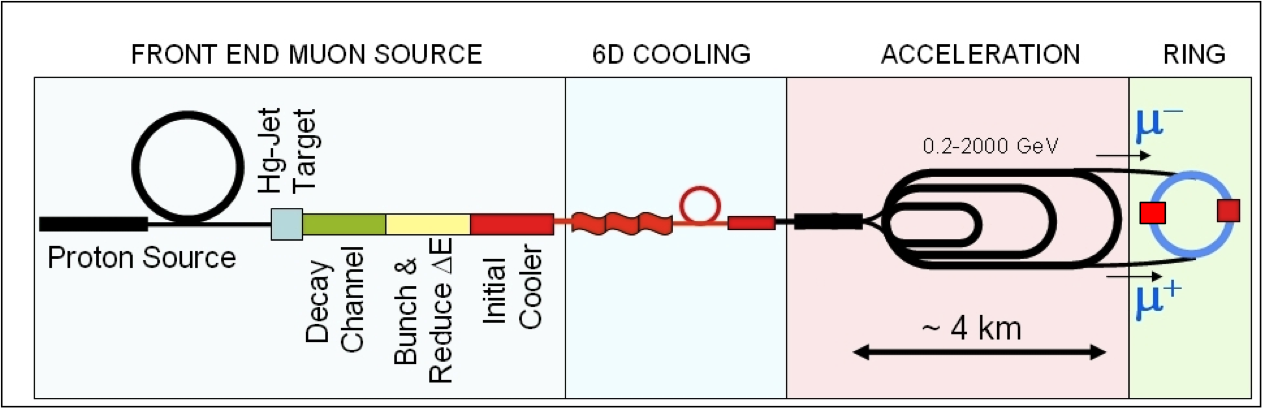}
\caption{Schematic of the muon collider complex..} \label{muon complex}
\end{figure*}

Figure  \ref{muon complex}  shows a schematic of a possible Muon Collider Complex.  The front-end muon source would 
use protons from a 4 MW accelerator (e.g. Project X) impinging on a mercury jet target.  This is followed by  
a decay channel, beam bunching and bunch rotation, and an initial cooling stage. This first stage could 
be identical to the front end of a Neutrino Factory.  The initial section is followed by a cooling section 
that would further cool the beam in both momentum and position space.   Finally the muons are 
accelerated in a multi-racetrack section before being injected into the storage ring. The Muon Collider 
accelerator is described in more detail in Mike Zisman's contribution to this conference\cite{Zisman} and in 
publications of the Muon Accelerator Program~\cite{Mucol}~\cite{Palmer:2007zzc}.

%%%%%%%%%%%%%%%%%%%%%%%%%%%%%%%%%%
\subsubsection{Ionization Cooling}
Previously used beam cooling techniques, such as stochastic or electron cooling will not cool a muon 
beam quickly enough for the muons to be used in a collider.  Instead, muons will be cooled utilizing 
ionization energy loss, combining a low Z absorber 
'with a high field solenoid and normally conducting RF cavity. Muons lose both transverse and longtitudinal 
momentum in the absorber.  The RF restores longtudinal momentum, reducing $p_{x,y}/p_z$.  Multiple 
scattering acts to reheat the muons, resulting in an equilibrium emittance that is proportional to $(1/(X_0 \times dE_\mu / ds)$. 
The optimum absorber material has maximum energy loss per radiation length.  Liquid hydrogen is the 
best candidate.
Muon cooling ideas are being tested in the MICE experiment at RAL. This experiment will measure the properties 
of  single muons  before and after passing through an ionization cooling section\cite{Karadzhov:2010zz}.

\subsubsection{RF Breakdown}
After the absorber longitudinal momentum of the muon is
restored by accelerating the beam through an rf cavity. The overall cooling effect 
is more efficient when a solenoid 
focuses the beam while still in the cavity. Unfortunately, the breakdown voltage of an RF cavity is 
significantly degraded in a magnetic field~\cite{Stratakis:2009zz}.  
Electrons are emitted from the cavity surface in regions which are not perfectly smooth. 
These electrons are guided in the magnetic field and impact on localized regions on the  
wall of the cavity.  In the absence of a field the electrons tend to be more dispersed and 
the resulting impact damage
 damage is not localized.  The MAP program is studying how these effects can be mitigated by improving the surface material properties 
(beryllium coatings) or by interrupting the free flow of the field-emitted electrons by filing the cavity with 
gas~\cite{Freemire:2011zz}.

\subsubsection{Neutrino Radiation}
Muons in the collider ring decay at the rate of $1.28 \times 10^{10} decays/m/s$.  At this rate the radiation
due to interacting neutrinos at the site boundary is a concern.  The neutrino interaction rate increases as $energy^3$.  
Local hot spots will occur which correspond to straight sections in the storage ring.  Off site radiation can be 
minimized by limiting the length of straight sections, increasing the depth of the collider ring, and managing 
the operating parameters to maximize luminosity/dose. However, neutrino-induced 
radiation may be the practical limit to the maximum energy of a machine located on the Fermilab site.

%\subsubsection{Beam Decay heat loads}
%Electrons from decay of the stored muon beam deposit 2.4 MW into the storage ring.  This energy has to be absorbed 
%while maintaining the superconducting magnets at low temperature.  Two approaches have been studied, an open 
%gap magnet that allows the energy to pass through the magnet, and a design with an inner tungsten shield.  It appears 
%that a design with a few cm radius water-cooled tungsten shield, followed by  superconducting  coils at the outer 
%radius, is preferred.  Trade-offs are being studied between absorber radius and cryogenic power needed to cool the outer 
%coils.  The current design uses less than half the tungsten cross section than the 1998 design, but with cryogenic wall power 
%increased from 1.8 to 10 MW.

\begin{table}[ht]
\begin{center}
\caption{Comparison of the machine parameters for a 1.5 and 3 TeV Muon Collider with designs for CLIC at 3 TeV.}
\begin{tabular}{|l|c|c|c|c|}
\hline & $\mu^+\mu^-$ & $\mu^+\mu^-$ & $e^+e^-$ CLIC & \\
\hline \textbf{CM Energy} & 1.5 & 3.0 & 3.0 & TeV \\
\hline \textbf{Luminosity} & 1 & 2-4 & 2 & $10^{34} cm^2 sec^{-1}$ \\
\hline \textbf{rms bunch height} & 6 & 4 & 0.001 & $\mu m$ \\
\hline \textbf{Diameter/length} & 2 & 4 & 48 & km \\
\hline \textbf{Wall power} & 147 & 159 & 560 & MW\\
\hline
\end{tabular}
\label{example_table}
\end{center}
\end{table}

%%%%%%%%%%%%%%%%%%%%%%%%%%%%%%%%%%
\section{Physics and Detector Studies}
Although there is a revived muon accelerator effort, there has been little corresponding detector effort which could 
study the large beam-related backgrounds with modern simulation tools and detector technologies.
This effort was commissioned at the Muon Collider 2011 meeting in Telluride at the end of June 2011. 
Initial studies were performed using ILCROOT, LCSIM, and GEANT frameworks, with the goal of understanding 
the characteristics of the backgrounds and what sorts of tools would be needed to build a detector 
for the Muon Collider environment. 

\begin{figure*}[ht]
\centering
\includegraphics[width=70mm]{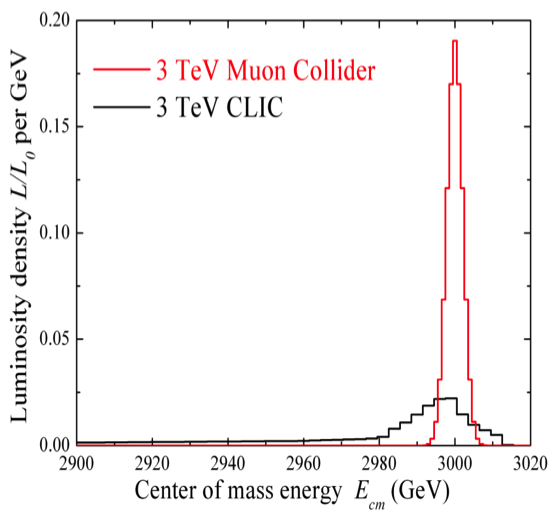}
\includegraphics[width=70mm]{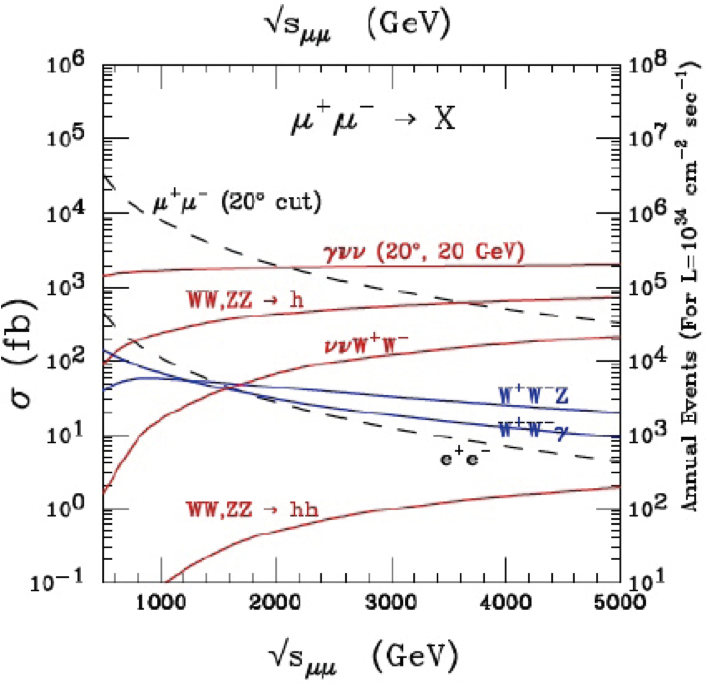}
\caption{(left) Beam energy distribution in an $e^+e^-$ and muon collider . The tail in the  $e^+e^-$ 
collider is due to beam-beam (beamstrahlung) effects and will occur in any $e^+e^-$ collider 
with a tight final focus. (right)Cross sections as a function on center-of-mass energy for various 
reactions in a muon collider. } \label{beamstrahlung}
\end{figure*}

Both CLIC and the Muon Collider would need to study new phenomena with a level of precision that 
can not be achieved at LHC.   The experimental conditions in CLIC and the Muon Collider are quite different. 
CLIC has a bunch train with 0.5 ns bunch spacing.  The Muon Collider will have a single bunch each of 
$\mu^+$ and $\mu^-$, colliding at $10 \mu s$ intervals.  Energy spread of the $e^+e^-$ colliding beams scales 
as $\delta E/E \propto \gamma^2$.  The effect of this on the center of mass energy resolution is shown in 
figure \ref{beamstrahlung}.  The excellent Muon Collider energy resolution allows for precise measurements of 
s-channel resonances and precise turn-on threshold scans. 
%At multi-tev energies both $e^+e^-$ and $\mu^+\mu^-$ collider cross sections become dominated by 
%fusion reactions, meaning that reactions such as $\mu^+\mu^- \rightarrow \nu_{\mu} \nu_{\mu} X$ become 
%important.  
The higher muon mass means that there is a significant rate for s-channel Higgs-like particle 
production which is not available in a $e^+e^-$ machine. Finally, although the muon beams are born 
polarized, the large phase space acceptance at low pion momentum needed for efficient muon collection 
means that the colliding beam polarization is likely to be below 20\%.  Additional polarization can be 
obtained at the expense of luminosity by limiting the initial phase space acceptance.

\subsection{Machine-Detector Interface and Backgrounds}
The physics environment of the Muon Collider is dominated by 
very large flux of high energy electrons from muon decay.  These electrons interact in the beam transport system as well as in the  shielding around the interaction point.  Carefully designed shielding, both for the experiment and the 
beam transport system is necessary to keep backgrounds manageable.
A feature that distinguishes the Muon 
Collider from other experiments is a tungsten/borated polyethylene ``nose" that extends at 
a  $10^{\circ}$ angle 6 cm from the interaction point. The cone is designed to absorb the intense electromagnetic 
radiation due to muon decay that accompanies the muon beams and reduces the overall background level by 
three orders of magnitude.
Over the past year there has been a substantial effort to fully model the backgrounds 
in a model detector at the Muon Collider.  There are now both MARS and G4beamline models of the 
interaction region with detailed simulations of the particle flux.  

\begin{figure*}[ht]
\centering
\includegraphics[width=135mm]{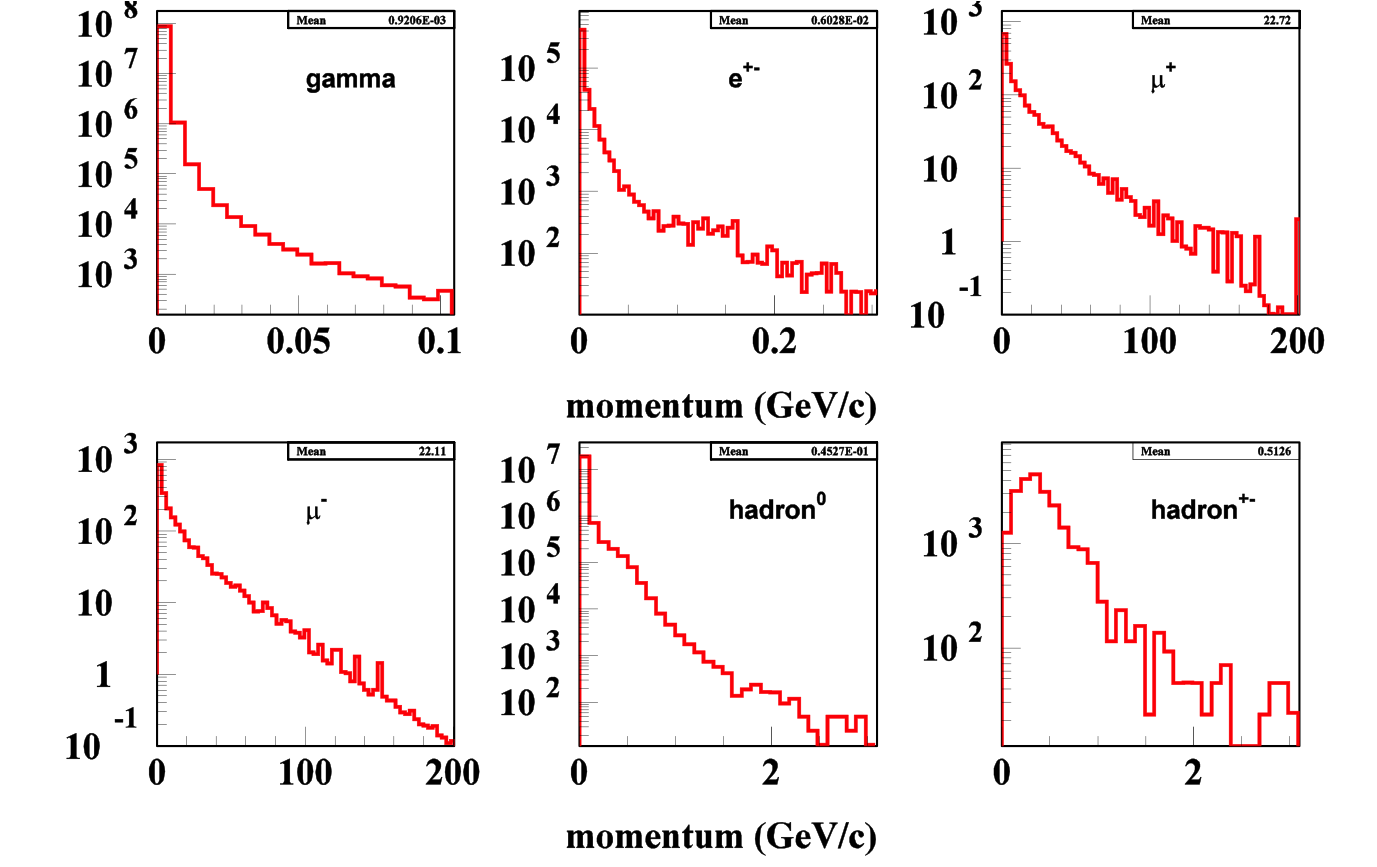}
\caption{Energy distributions of particles entering the detector region from a MARS simulation 
of Muon Collider beam backgrounds\cite{Striganov}.} \label{background1}
\end{figure*}

Figure \ref{background1} shows the background flux entering the detector region in a typical 
Muon Collider interaction.  Total non-ionizing background is about 10\% that of the LHC, 
but the crossing interval is 400 times longer, resulting in high instantaneous flux.  
The background is very different in character than that of either the LHC or CLIC.
It is dominated by soft photons and low energy neutrons emerging from the shielding surrounding the 
detector.  A typical background event has 164 TeV of photons, 
172 TeV of neutrons, and 184 TeV of muons. With the exception of muons and charged hadrons the 
background spectrum is dominated by low energy particles.
Only a small fraction of the background originates from the vicinity of the interaction region.  
This means that most of the decay background 
is out of time with respect to particles originating from the $\mu^+\mu^-$ collision.

\subsection{Background Rejection Techniques}
The fact that much of the background is soft and out of time gives us two handles on the design 
of an experiment that can cope with the high levels of background.  Timing is especially 
powerful.  Figure~\ref{background1} shows the fraction of 
background and signal hits in the tracker preserved as a function of the width of the timing gate.   An important 
feature is that the local gate t=0 is defined as the time when relativistic particle emerging from the interaction 
point arrives at the detector.  Therefore a very tight cut can be made, still preserving the bulk of the tracks of interest. 
A 1 ns cut rejects two orders of magnitude of the overall background and about 4 orders of magnitude of 
neutron background. 

A detailed tracking study was performed using the ILCROOT framework which includes a 
all silicon detector similar to the one proposed for SiD\cite{Mazzacane}.  The tracker is fully pixelated with $50 \mu m$
pixels in the tracker and $20 \mu m$ pixels in the vertex section.  Hits are required to have a 
threshold of 3000 electrons, which eliminates much of the soft photon and neutron background. 
Reconstructed tracks are required to have an impact parameter less then 3 mm. With no timing cut 
the reconstruction program fails due to the large number of hits. 
However if the $\Delta t$ cut is 3 ns only 11 background tracks are found 
and a 1 ns cut further reduces this to 3 tracks, all with low momentum.  Replacing the timing cut 
by a time stamp would allow the hit arrival time to be used in the track fit, providing both background 
rejection and some level of particle identification.

Soft, uncorrelated hits can also be eliminated by exploiting the correlation between two closely 
spaced tracking layers.  This was first suggested by Steve Geer in early Muon Collider studies 
and a similar concept is being explored for a track trigger for CMS\cite{Lipton:2011zz}. In that 
design two silicon sensors are spaced by a 1 mm thick interposer and only hits which are 
correlated between the top and bottom layers are used for the trigger.  Initial 
studies indicate that such an arrangement would also be extremely effective at reducing 
Muon Collider backgrounds and might reduce dependence on the timing cut.  

\begin{figure*}[ht]
\centering
\includegraphics[width=100mm]{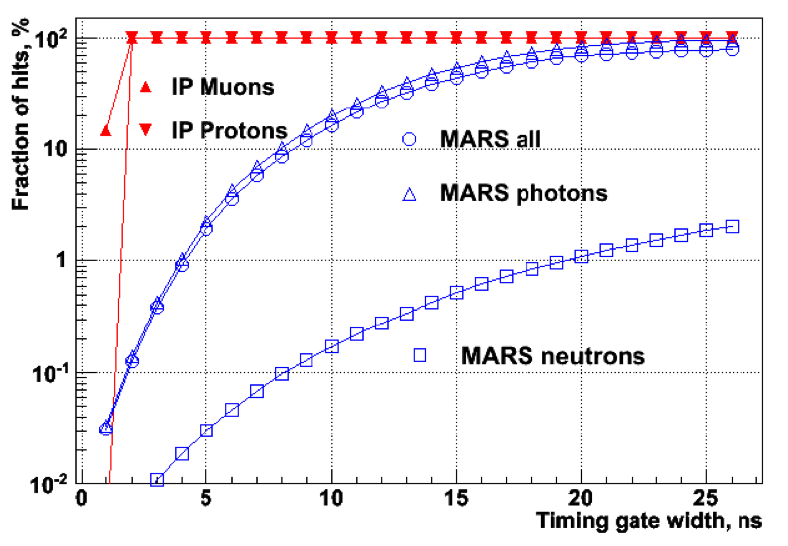}
\caption{Fraction of hits accepted as a function of the timing gate width\cite{Terentev}. The gate  
start is defined by the time of flight for a relativistic particle emerging from the 
interaction point. } \label{track_timing}
\end{figure*}

Timing is also crucial for calorimetry.  Our initial ILCROOT simulation studied a dual 
readout  ``Adriano" heavy glass/scintillator calorimeter with  $4 \times 4$ cm cells with 7.5 interaction lengths~\cite{Vito}. 
There are two longitudinal sections 20 and 160 long, each with front and rear readout through 
SIPMs.  Different timing cuts are used for the scintillation and Cerenkov light in the front section (15 and 6 ns, 
respectively) with a 22 ns cut for the rear section.  
In the central barrel region backgrounds deposit an average energy per tower of 5.33 GeV per event with 
RMS fluctuations of 540 MeV in the front section.  The rear section sees an average energy of 630 MeV with 
fluctuations of 430 MeV.  Further analysis will be needed to optimize cuts and determine jet energy resolution.

An optimal calorimeter design might combine fast timing with the reconstruction ability of pixelated 
calorimeters being studied for particle flow.  A pixelated  imaging sampling calorimeter with 
200 $\mu m$ square cells was proposed by R. Raja~\cite{Raja}. In this design a 2 ns  ``traveling trigger" gate 
referenced to the time of flight with respect to the 
beam crossing is used to reject out-of-time hits.  In this case background 
rejection was found to be  $3 \times 10^{-2}$ to $4 \times 10^{-4}$.  This sort of calorimeter can implement 
compensation by recognizing  hadronic interaction vertices and using the number of such 
vertices to correct the energy.  Initial estimates of the resolution of such a compensated calorimeter 
is $60\%/\sqrt{E}$. In contrast to relativistic tracks and electromagnetic showers, hadronic showers
can take significant time to develop\cite{simon}. Further study is needed to understand the tradeoff between background 
rejection provided by a short time gate and the loss of energy resolution 
caused by the slow time development of hadronic showers .  

We have learned that tracking seems possible in a Muon Collider detector.
Calorimetery is more challenging, but progress is being made on imaging calorimeter concepts
that appear to meet the physics needs.
Precise timing and pixelated detectors will be crucial to a successful Muon Collider detector.  
Both come at a cost.  For example the time resolution, $\sigma(t) \approx rise time \times (noise/signal)$.
The signal/noise and gain-bandwidth of typical electronic front ends are proportional to the transductance 
of the front end transistor - which in turn is proportional to front-end current.  This means that 
electronics will necessarily dissipate significant power and,
in contrast to planned ILC detectors, detectors for the Muon Collider will have to be water cooled 
with associated increase in mass.  The large background of non-ionizing radiation means that 
silicon detector will have to be kept cold, around -10 C, again increasing the detector mass.  Such 
a detector will resemble an LHC experiment more closely than those planned for ILC or CLIC.

%%%%%%%%%%%%%%%%%%%%%%%%%%%%%%%%%%
\section{Conclusions}
Both accelerator and detector aspects of a muon collider are extremely challenging.  The Muon Accelerator 
Program has been formed to study and evaluate the accelerator challenges.  A complementary effort in now 
beginning to study physics and detector aspects.  Initial results of these studies indicate that detectors can be 
designed that withstand the fierce backgrounds.  Such a detector is likely to be more massive than a 
corresponding ILC detector, but could have unique capabilities.  Fast timing and fine segmentation appear to 
be crucial.

%\begin{acknowledgments}

%\end{acknowledgments}

\bigskip % extra skip inserted

\end{document}